
%
\magnification=\magstep1
\parindent=1pc
\baselineskip=24pt
\parskip 2pt plus 1pt
\def\[#1]{$^{#1}$}       
\centerline{\bf Magnetic impurities coupled to quantum antiferromagnets}
\centerline{\bf in one dimension}
\centerline{J. Igarashi}
\centerline{\it Faculty of Engineering, Gunma University,
Kiryu, Gunma 376, Japan}
\centerline{T. Tonegawa}
\centerline{\it Faculty of Science, Kobe University,
Rokkodai, Kobe 657, Japan}
\centerline{M. Kaburagi}
\centerline{\it Faculty of Cross-Cultural Studies, Kobe University,
Rokkodai, Kobe 657, Japan}
\centerline{P. Fulde}
\centerline{\it Max-Planck-Institut f\"ur Physik komplexer Systeme,}
\centerline{\it Bayreuther Str. 40 Hs. 16, D-01187 Dresden, Germany}
\smallskip
\centerline{(Received \hskip 3cm)}
\smallskip
\leftline{Abstract}
Magnetic impurities coupled antiferromagnetically to a one-dimensional
Heisenberg model are studied by numerical diagonalization of chains of
finite clusters.
By calculating the binding energy and the correlation function,
it is shown that a local singlet develops around each impurity.
This holds true for systems with a single impurity,
with two impurities, and for impurities forming a lattice.
The local character of the singlet is found to be little affected
by the presence of other impurity spins.
A small effective interaction is found between a pair of impurity spins,
which oscillates depending on impurity distances.
For impurity lattices, the energy spectrum shows a gap which is found
to be much smaller
than the binding energy per impurity if the coupling constants are small.
For larger coupling constants, it increases to the same order of magnitude
as the binding energy, indicating that a local singlet is broken
to create excited states.
Impurity lattices with ferromagnetic couplings are also studied
and their connection to the Haldane problem is discussed.
\bigskip
\leftline{PACS numbers: 75.10Jm, 75.20.Hr, 75.30.Hx, 75.30.Mb}
\vfil \eject
\centerline{\bf I. INTRODUCTION}
\smallskip
A magnetic impurity embedded in a system of conduction electrons
forms at zero temperature a spin singlet
through the antiferromagnetic Kondo coupling.
This Kondo problem has been extensively studied and even
solved exactly.\[1,2]
In these studies, the interaction between conduction
electrons is usually neglected. The general question arises what changes
we should expect when the interaction between conduction electrons becomes
important so that they are strongly correlated.
As a first step toward an answer of this question, we consider
strongly interacting conduction electrons in a half-filling band in one
dimension. Charge fluctuations are suppressed by the strong interaction,
and the physics of the low-lying excitations is well described
by the spin-${1\over 2}$ antiferromagnetic Heisenberg model in this strong
coupling limit. The Hamiltonian is given by\[3]
$$
 H = H_c + H_{fc}, \eqno(1.1)
$$
where
$$
\eqalignno{
 H_c &= J\sum_{<i,j>}\bigl[ S_{ic}^z S_{jc}^z +\alpha(
 S_{ic}^x S_{jc}^x+S_{ic}^y S_{jc}^y)\bigr],\quad \alpha=1, &(1.2)\cr
 H_{fc}&= J'{\bf S}_{0c}\cdot{\bf S}_{0f}, &(1.3)\cr}
$$
with $J>0$, $J'>0$.
Here ${\bf S}_{ic}$ and ${\bf S}_{0f}$
denote the spin-${1\over 2}$ operators for the electron at site $i$
of the chain and for an electron in an impurity which is close to site 0.
The notation $<i,j>$ refers to nearest-neighbor sites in the chain.
A periodic boundary condition is imposed on the chain system.
The system is schematically shown in Fig.1.

In the presence of an Ising-like anisotropy ($\alpha<1$ in Eq.(1.2)),
the ground state is N\'eel ordered, and the excitations described by $H_c$
have an energy gap.
In the classical picture, the impurity spin is antiparallel to the neighboring
spin in the chain. Quantum fluctuations
reduce $|<S_{0f}^z>|$ from its classical value, but do not completely
destroy the classical picture. [Here $<A>$ denotes the average of $A$
with respect to the ground state.] As analyzed in the Appendix on the basis
of a spin-wave expansion,\[4,5] $|<S_{0f}^z>|$ decreases rapidly when
$\alpha$ approaches 1, while the transverse spin correlations
$|<S_{0f}^xS_{0c}^x>|$ grow to nearly maximum value 0.25.
Although the spin-wave expansion may not work well for $\alpha\to 1$,
these results suggest that in this limit the quantum fluctuations are large
enough for the impurity spin to form a singlet with the spins of the chain.
The purpose of this paper is to study the nature of this singlet formation
for $\alpha=1$ by exact diagonalization of finite-size clusters,
which has been successful to study quantum spin models
in one dimension.\[6]
In order to obtain a comprehensive understanding of the problem,
we consider not only the single impurity case but also the cases when
two impurities are present or when the impurities form a lattice (see Fig.1).

If one of the impurity spins forms a singlet with the spin in the chain
closest to the impurity site, the energy gain by $H_{fc}$ is
${3\over 4}J'$.
This local singlet, however, will lose some energy due to the presence of
$H_c$.
Calculating the binding energy, we find that the energy gain per impurity spin
is considerably reduced from ${3\over 4}J'$.
As shown later, the reduced values are, however, still much larger than those
for the conventional Kondo problem,\[7] suggesting that the binding energy is
enhanced by increasing interactions between conduction electrons.
It is also found that the binding energy per impurity spin is nearly
independent of the number of impurities.

We also calculate the correlation function between an impurity spin
and the spin in the chain closest to the impurity site.
The absolute value is found to increase rapidly with increasing $J'$.
Just like the binding energy, it is nearly the same for the three different
systems under consideration. Even for small values of $J'$, this coincidence
seems to hold.
The local disturbance generated by an impurity site seems to be little
influenced by the presence of other impurities.
However, this does not necessarily mean that
correlations between impurities are small; as shown later, they may be
substantial, particularly for small values of $J'$.
The reason is that a local triplet is admixed to the singlet state
due to interaction with the other spins in the chain
and this results in an indirect interaction between impurities
and an increase in the binding energy.
In the conventional Kondo model, the two-impurity problem has been
studied by several methods.\[8-10]
Thereby a main issue is the interplay between the single-site Kondo effect and
the RKKY interaction.
In particular, the half-filled one-dimensional Kondo lattice
has been studied by several authors.\[11-13]
It is known that for small Kondo-coupling constants
the RKKY interaction becomes more important than
the Kondo-singlet formation.\[13]
The present findings extend those observations to the case of strongly
correlated electrons.

In the impurity-lattice case, we also calculate the energy gap between
the lowest excited states and the ground state. It is found that
they are triplets with momentum $k=\pi$
when the lattice constant is set equal to one.
The gap is very small for small values of $J'$. This is consistent with a
non-analytic dependence
$\propto{\rm e}^{-a/J'}$ where $a$ is some constant which is found for
the half-filled one-dimensional Kondo lattice.\[11] The gap increases
with increasing values of $J'$,
and eventually becomes of the order of the binding energy per impurity spin.
This suggests that the excited states are local triplets which are
created by breaking up local singlets.
This result is consistent with the one found for
the double chain model,\[14] which is slightly different from the present
model.

In addition to the studies involving an antiferromagnetic coupling,
we also study systems of impurity lattices with a ferromagnetic coupling
($J'<0$) to the spins in the chain.
By calculating the binding energy and the correlation function,
it is shown that local triplet states develop with increasing values
of $|J'|$. As a whole, the ground state is a singlet, and
the lowest excited states are triplets with momentum $k=\pi$, just as before.
The energy gap is very small, and increases gradually
with increasing values of $|J'|$. It is much smaller than the
binding energy per impurity spin,
indicating that the excited states do not destroy
the local triplets which are mainly responsible for the binding energy.
In the limit $J'\to -\infty$, the gap goes over into the
Haldane gap.\[15] In this context, there have been several models studied
which are slightly different from the present one, such as the
ferromagnetic-antiferromagnetic alternating Heisenberg chain model\[16]
or the double chain model.\[17]
The present results are consistent with the ones found previously.

In Sec.II, numerical analysis is given of exact diagonalization
of finite clusters.
Section III contains the concluding remarks.
In the Appendix, a system is analyzed within the spin-wave expansion,
in which an impurity spin is coupled to an Ising-like
anisotropic chain.
\bigskip
\centerline{\bf II. NUMERICAL ANALYSIS}
\smallskip
In the following we present the results for binding energies
as well as various pair correlation functions of finite Heisenberg chain
systems with a single impurity, with two impurities,
and with impurities forming a lattice.\[18] Thereby
we measure energies in units of J.
\bigskip
\centerline{\bf A. Single impurity}
\smallskip
We consider the Hamiltonian given by Eq.(1.1) with $\alpha=1$.
The ground state is a singlet.
The binding energy $\Delta E_1(J',N)$ is defined by
$$
 \Delta E_1(J',N)=E_g(0,N)-E_g(J',N), \eqno(2.1)
$$
where $E_g(J',N)$ denotes the ground-state energy for a spin chain
consisting of $N$ sites with an impurity coupled to it according to $H_{fc}$.
The number of sites $N$ varies from 4 to 23.
Values for $\Delta E_1(J',N)$ are naturally grouped into two series
according to even or odd numbers of N.
The values for odd numbers N are larger than those for even numbers.
This is explained as follows. The ground state for chains with
odd numbers N is a doublet. From it a localized doublet may be easily
generated to form a singlet with the impurity spin. But the ground state
for chains of even numbers N is a singlet, from which a localized
doublet cannot be created.
With increasing N, binding energies decrease for N being odd,
while they increase for even numbers of N.
Since effects caused by the impurity spin are expected to be confined
to a finite range from the impurity site, one might think that
the binding energy is independent of the system size if it
exceeds the size of the spin correlation. This is not the case though,
since the energy gain due to singlet formation depends on the energy
spectrum of the chain, which is discrete and depends on the system size.
When extrapolating each series to $N\to\infty$, we assume a dependence of
$\Delta E_1$ in powers of $1/N$,
$$
 \Delta E_1(J',N)=\Delta E_1(J') + {{a_1}\over N}
 + {{a_2}\over{N^2}} + {{a_3}\over{N^3}} + \cdots, \eqno(2.2)
$$
to which we fit the data by a least-square method.\[19]
Since the two series approach each other,
we can evaluate accurately their limiting value.
If the impurity spin forms a perfect singlet with the spin in the chain
closest to it, the energy gain due to $H_{fc}$
would be ${3\over 4}J'$.
This singlet state suffers an energy loss through $H_c$ though.
Therefore some compromise between the two energies must take place.
We find that the net binding energy is considerably reduced from
${3\over 4}J'$, with a factor which is larger for smaller values of $J'$.

We also calculate the correlation function
$<{\bf S}_{0f}\cdot{\bf S}_{0c}>$
between the impurity spin and the spin in the chain closest to the impurity
site. Figure 3 shows calculated values of $-<{\bf S}_{0f}\cdot{\bf S}_{0c}>$
as function of $J'$.
Like for the binding energy, they fall into two series
according to odd and even values of N. The correlation is antiferromagnetic.
With increasing N, $-<{\bf S}_{0f}\cdot{\bf S}_{0c}>$ decreases
for odd values of N,
while it increases for even values, just like in the case of the binding
energy.
We extrapolate each series to $N\to\infty$ in a similar way as done
for the binding energy, although the assumption of a power expansion in
$1/N$ is uncertain for this quantity.
The values extrapolated by the two
series are so close that we obtain reliable
estimates of the correlation function for $J'>0.2$.
For small values of $J'$ ($<0.2$), however, the two series do not approach
each other closely enough to obtain reliable estimates.
This suggests that the assumption concerning the power expansion
in $1/N$ is inadequate.
The absolute values first increase rapidly and then move gradually toward
0.75 with increasing values of $J'$, which is the value for a perfect singlet.
Contributions from triplet states of the two spins ${\bf S}_{0f}$ and
${\bf S}_{0c}$ appreciably influence the value of
$<{\bf S}_{0f}\cdot{\bf S}_{0c}>$ provided $J'$ is not too large.
Another way of stating this is by saying that the impurity spin forms
a singlet with the spins of the chain over a region which is rather extended.
Spin correlations between an impurity spin and that of conduction electrons
in a Hubbard chain have also been considered by Hallberg and Balseiro,\[20]
who found an oscillatory behavior with distance.
\bigskip
\centerline{\bf B. Two impurities}
\smallskip
In this subsection, we study the case of two impurities placed
on nearest-neighbor and on next-nearest-neighbor sites, respectively,
as shown in Fig.1(b).

The binding energies, $\Delta E_2^{\rm nn}(J',N)$, $\Delta E_2^{\rm nnn}(J',N)$
are defined by the same equation as Eq.(2.1),
where superscripts nn and nnn stand for
nearest-neighbor and next-nearest-neighbor positions,
respectively.
Figure 4 shows the binding energy per impurity spin
$\Delta E_2^{\rm nn}(J',N)/2$.
The series of values for even and odd numbers of sites $N$ approach each other
more rapidly than for the single-impurity case.
We have also calculated\break
$\Delta E_2^{\rm nnn}(J',N)/2$, whose dependence
on $J'$ and $N$ is similar to $\Delta E_2^{\rm nn}(J',N)/2$.
Figure 5 shows the extrapolated values to $N\to\infty$
for $\Delta E_2^{\rm nn}(J')/2$, $\Delta E_2^{\rm nnn}(J')/2$
and their comparison with $\Delta E_1(J')$.
Both values are close to $\Delta E_1(J')$, indicating that
the presence of another impurity spin affects the first one so little
that the impurities
can be considered as being almost independent.
The value $\Delta E_2^{\rm nn}(J')/2$ is slightly larger, while
$\Delta E_2^{\rm nnn}(J')/2$ is slightly smaller than $\Delta E_1(J')$.
The mutual influence of impurities, though it is small, seems to show
a small energy gain or loss respectively, i.e., an oscillatory behavior,
depending on the distance between the two impurities.

Figure 6 shows the correlation function $-<{\bf S}_{0f}\cdot{\bf S}_{0c}>$
between one of the impurity spins and the spin in the chain closest to it,
when the two impurities are placed onto nearest-neighbor sites.
A local singlet develops with increasing values of $J'$.
By comparing Fig.6 with Fig.3, we find that
the extrapolated values are very close to those for the single-impurity case.
Therefore the presence of second impurity has little effect on the
pair correlation, in agreement with the findings for the binding energy.

The small mutual effect on both, the binding energy and the correlation
function $<{\bf S}_{0f}\cdot{\bf S}_{0c}>$, suggests a small
correlation $<{\bf S}_{0f}\cdot{\bf S}_{1f}>$ between two impurity spins.
Figure 7 shows the calculated values of $-<{\bf S}_{0f}\cdot{\bf S}_{1f}>$
for impurities placed on nearest-neighbor sites.
Against what we would expect, their absolute values are not small.
An impurity spin does not form a perfect singlet with
the spin in the chain closest to it, instead a triplet state
is mixed in, as shown in Fig.6. We speculate that
an indirect interaction, though small, is based on these
degrees of freedom, thereby affecting considerably the spin correlation.
The correlations decrease monotonically with increasing values of $J'$
for even numbers N, while they first increase and then decrease for odd
numbers N.
Their decrease with increasing $J'$ ($J'>0.25$) may be taken
as an indication of the development of a singlet state between each impurity
spin and the spins in the chain,
resulting in weaker correlations between impurity spins.
The values extrapolated to $N\to\infty$ for even and odd values of N are
sufficiently close in order to allow for reliable estimates of the limiting
behavior provided $J'>0.25$.
But when $J'\to 0$, the values for even and odd numbers of N
differ considerably.
This may be interpreted as follows.
In the absence of $H_{fc}$, the singlet and triplet states of two impurity
spins are degenerate. Now let us turn on $H_{fc}$.
If $J'$ is smaller than the finite gap in the excitation energy of
the finite-size systems, the state of the spins in the chain is little
modified by the impurity spins.
For even values of N, the ground state of the spins in the chain is `nearly'
a singlet.
Since the total system must be a singlet, the singlet state between
two impurity spins is favored,
resulting in large absolute values of the pair correlation.
On the other hand, for odd values of N, the ground state of the spins
in the chain is `nearly' a doublet, and that of the total system must be
a doublet.
Both, the singlet and triplet states of the two impurity spins, can
couple to the spins in the chain to form a doublet.
In fact, for sufficiently large systems in which the gap is
much smaller than $J'$, the states of the spins in the chain can easily
be deformed locally to form a singlet with the impurity spins.
It seems difficult to obtain reliable limiting values for the correlation
function from the present sizes of chain when $J'$ is small.
\bigskip
\centerline{\bf C. Impurity lattice}
\smallskip
In this subsection, we study systems in which the impurities form a lattice,
as shown in Fig.1(c).
Each impurity spin is coupled to a spin in the chain with exchange
constant $J'$, and the impurities do not interact directly with each other.

We calculate the binding energy per impurity spin
which is defined by
$$
 E_B^{\rm lat}(J',N) = \left[E_g(0,N)-E_g(J',N)\right]/N. \eqno(2.3)
$$
Figure 8 shows the calculated values for $E_B^{\rm lat}(J',N)$.
Although the sizes of chain are smaller ($N=3\sim 14$) than for the single
impurity case, the two series for even and odd values of N approach each other
very closely, and therefore we can extrapolate them accurately to $N\to\infty$.
The extrapolated value of $E_B^{\rm lat}(J')$ is nearly the same as for
$\Delta E_1(J')$.

We also calculate the correlation function
between one of the impurity spins and the nearest spin in the chain.
Figure 9 shows the calculated function $-<{\bf S}_{0f}\cdot{\bf S}_{0c}>$.
For this quantity, it is difficult to extrapolate the value to $N\to\infty$,
partly because of small sizes of chain. But even without
the extrapolation, we can guess rather accurately the limiting
value, since the two series for even and odd values of N are already
sufficiently close. The amplitude of the
singlet increases with increasing values of $J'$.
This value is nearly the same as for the single-impurity case.
When taken together with the result for the binding energy, we may state that
the local character of each impurity is little influenced by the presence of
the impurity lattice.

Inspecting $E_B^{\rm lat}(J')$ more closely, we notice that it is slightly
larger than $\Delta E_2^{\rm nn}(J')/2$, as shown in Fig.5.
Forming an impurity lattice leads to an increase in the binding energy.
Also comparing $<{\bf S}_{0f}\cdot{\bf S}_{0c}>$ with that
for the single impurity case shown in Fig.3, we notice
that the absolute values are slightly larger here.
Forming an impurity lattice results in an increase in the amplitude
of the local singlet.

Figure 10 shows the correlation function
$-<{\bf S}_{0f}\cdot{\bf S}_{1f}>$
between a pair of impurity spins on nearest-neighbor sites.
The dependences on $J'$ and $N$ look complicated,
particularly for small values of $J'$.
The extrapolated values first increase and then decrease
with increasing values of $J'$. For $J'>0.5$, the values are very close to
those
for the two-impurity case (Fig.7),
indicating again that forming a lattice affects little the pair correlations
between impurities.

In addition to the ground-state properties, we also calculate the energy
and the wave function of the lowest excited states.
The lowest excited states are triplets with momentum $k=\pi$.
The energy gap $\Delta(J',N)$ is defined by the difference
between the energy of the lowest excited state and of the ground state.
Figure 11 shows the calculated values of $\Delta(J',N)$.
The gap decreases with increasing $N$ (even and odd),
and no lower limit of the gap exists.
The extrapolated value $\Delta(J')$ is much smaller than $E_B^{\rm lat}(J')$
for small values of $J'$ ($<0.5$). For $J'<0.3$ it is nearly zero.
This favors a non-analytic dependence on $J'$, i.e.,
$\Delta(J')\propto {\rm e}^{-a/J'}$ with
$a$ some numerical constant, as known for Kondo lattices.\[11]
Clearly we cannot rule out other dependences on $J'$
or the presence of a critical value of $J'$ below which the system is gapless.
For larger values of $J'$ ($>0.7$), the gap is not much different from
$E_B^{\rm lat}(J')$.
This means that around each impurity a local singlet is well developed,
and that it is broken up when an excited state is generated.
This process requires an energy of order $E_B^{\rm lat}(J')$.

Breaking up a local singlet to create excited states
becomes apparent in the strong coupling limit when $J'\to\infty$.
In that case, each impurity spin forms a perfect singlet with the nearest spin
in the chain. The low-lying excited states consist of
a triplet excitation at a particular site
with the other sites remaining within the singlet state.
They are 3N-fold degenerate, having energy $J'$.
The first-order correction to order $1/J'$ lifts the degeneracy.
Let $|T^\mu(j)>$ be the state of the triplet to which site $j$ is excited
with the magnetic quantum number $\mu$($=\pm 1,0$).
Then the state $|\psi^\mu(k)>$
$= \left({1\over N}\right)^{1/2}\sum_j {\rm e}^{-ikj}|T^\mu(j)>$ is an
eigenstate with an excitation energy $E_k=J'+{1\over 2}\cos k$.\[14]
Note that $E_k$ has a minimum at $k=\pi$, which is
consistent with the present numerical results for weak and intermediate
couplings.

Finally, we discuss briefly the changes which occur when the
coupling is ferromagnetic $J'<0$. The ground state is again a singlet.
Figures 12 shows the binding energies per impurity spin when the impurity
number changes.
Just like in the case of $J'>0$, results fall into two groups
corresponding to even and odd values of N.
If each impurity spin forms a perfect triplet with the nearest spin
of the chain, the energy gain due to $H_{fc}$ is ${1\over 4}|J'|$.
The extrapolated values are again smaller than ${1\over 4}|J'|$, but
the reduction ratio is less than for $J'>0$, as is noticed by comparison
with Fig.8.

Figure 13 shows the correlation function $<{\bf S}_{0f}\cdot{\bf S}_{0c}>$
between one impurity spin and the nearest spin in the chain as $J'$ varies.
The values first increase rapidly and then move gradually toward 0.25 ,
which is the one of a perfect triplet.
By comparison with Fig.9, the local triplet character seems to be established
rapidly.
Similarly as in Fig.9, it is difficult to extrapolate the value for
$N\to\infty$.

The lowest excited states are found to be triplets with momentum $k=\pi$.
Figure 14 shows the energy gap $\Delta(J',N)$ between these states.
The extrapolated values are very small;
for $-J'<0.2$, the gap is nearly zero.
They increase very slowly with increasing values of $-J'$.
The gap is much smaller than $E_B^{\rm lat}(J')$ given in Fig.12,
indicating that the lowest excited states can be formed
without destroying the local triplets.
In the limit $J'\to -\infty$, each impurity spin forms
a perfect triplet with the nearest spin in the chain.
Therefore we can regard each site as
being occupied by a spin $S=1$ with effective exchange coupling constant
$J_{\rm eff}=1/4$. For such a system it is known that the lowest excited state
has momentum $k=\pi$, and a gap which is given according to Haldane by
$\Delta(J')\simeq 0.411J_{\rm eff}=0.103$.\[21]
Although for $J'\sim -1.0$ the system is to good approximation
locally in a triplet state,
the gap is found to be still much smaller than the Haldane gap.
\bigskip
\centerline{\bf III. CONCLUDING REMARKS}
\smallskip
We have studied systems of magnetic impurities coupled to a spin-${1\over 2}$
antiferromagnetic Heisenberg chain.
We have calculated the binding energy per impurity spin and the
correlation function between an impurity spin and the nearest spin
in the chain for systems with a single impurity,
with two impurities, and with impurities forming a lattice.
We have shown that the local singlet character increases with
increasing antiferromagnetic coupling.

In the conventional Kondo problem, the binding energy may be written as
$\Delta E_1 \simeq (3\ln 2t)(\rho J')^2t + 4t{\rm e}^{-1/\rho J'}$
for weak couplings,\[22]
where $4t$ is the band width and $\rho=(2\pi t)^{-1}$ is the density of
states of the conduction electrons at the Fermi energy.\[7]
The first term is the normal part, and the second term is the anomalous part
corresponding to the Kondo temperature.
For $t\simeq 1$eV and $J'\simeq 0.2$eV, we obtain
$\Delta E_1\simeq 0.002$eV$+4{\rm e}^{-10\pi}$eV, i.e.,
the anomalous part is extremely small.
On the other hand, in the present Heisenberg chain model, the exchange
interaction between conduction electrons may be estimated as
$J(=4t^2/U)\simeq 1$eV when $U\simeq 4$eV and $t\simeq 1$eV,
where $U$ is the Coulomb interaction between conduction electrons.\[7]
For $J'\simeq 0.2$eV, we have $J'/J=0.2$, and from Fig.2 we estimate
the binding energy as $\Delta E_1\simeq 0.025$eV. This value is
at least by one order of magnitude larger than that for the conventional
Kondo problem.
This enhancement results from the fact that the disturbance in the chain
due to the presence of the impurity is much more local for finite $U$
than for $U=0$.

We have found that the binding energy per impurity spin as well as various
spin correlation functions are nearly independent of the impurity density.
This implies that the state which an impurity is forming with its
surroundings remains unaffected by the presence of other impurities,
which indicates its local character.
The binding energy is mostly determined by this local character.
Following the RKKY interaction, we may express the indirect interaction
between impurities at sites $i$ and $j$ as
$H_{\rm ind}=J'^2\chi(i,j){\bf S}_{if}\cdot{\bf S}_{jf}$,
where $\chi(i,j)=-i\int_0^\infty <[S_{jc}^z(t),S_{ic}^z]>
{\rm e}^{-\delta t}{\rm d}t$ with $\delta\to 0^+$.
This interaction was estimated to be small here.
Although the interaction energy is small, the spin correlations
$<{\bf S}_{if}\cdot{\bf S}_{jf}>$ are considerable for small values of $J'$.
They decrease though with increasing values of $J'$.
For large $J'$ each impurity spin is forming a singlet with the nearest spin
of the chain.

Clarke, Giamarchi, and Shraiman\[23] have recently studied impurity spin
models by transforming them
with the help of the bosonization technique
to the two-channel Kondo model.
They have found an anomalous temperature
dependence for the susceptibility.
Numerical studies of such a finite-temperature behavior are left for the
future.
\bigskip
\centerline{\bf ACKNOWLEDGMENTS}
\smallskip
One of the authors (J.I.) would like to thank MPI for Physics of Complex
Systems for financial support.
This work was partially supported by a Grant-in-Aid for Scientific Research
on Priority Areas, ``Computational Physics as a New Frontier in Condensed
Matter Research", from the Ministry of Education, Science and Culture, Japan.
\bigskip
\centerline{\bf APPENDIX}
\smallskip
We analyze the case that an impurity is coupled to an anisotropic Heisenberg
model by using the spin-wave expansion.
For an Ising-like anisotropy ($\alpha<1$ in Eq.(1.2)),
the ground state is N\'eel ordered, and
the system is divided into A (up spins) and B (down spins) sublattices.
In the following, the impurity spin is assumed to be antiferromagnetically
coupled to a spin in sublattice A, thus pointing downward.
We treat the deviation from the classical directions
by using the Holstein-Primakoff transformation:
$$
\eqalignno{
 S_{ic}^z &= S - a_i^\dagger a_i ,\quad
 S_{ic}^+ = (S_{ic}^-)^\dagger \simeq \sqrt{2S}a_i ,&(A1)\cr
 S_{jc}^z &= -S + b_j^\dagger b_j ,\quad
 S_{jc}^+ = (S_{jc}^-)^\dagger \simeq \sqrt{2S}b_j^\dagger ,&(A2)\cr
 S_{0f}^z &= -S + c^\dagger c ,\quad
 S_{0f}^+ = (S_{0f}^-)^\dagger \simeq \sqrt{2S}c^\dagger .&(A3)\cr}
$$
Here $a_i$ and $b_j$ are boson annihilation operators for spins on
sublattices A and B, respectively, and c is a boson annihilation operator for
the impurity spin. Substituting Eqs.(A1)-(A3) into Eqs.(1.2) and
(1.3), we have
$$
\eqalignno{
 H_c &= -JS^2N + JS\sum_{<ij>}\bigl[
  a_i^\dagger a_i + b_j^\dagger b_j + \alpha(a_ib_j+a_i^\dagger b_j^\dagger)
 \bigr], &(A4)\cr
 H_{fc} &= -J'S^2 + J'S(a_0^\dagger a_0 + c^\dagger c
  + a_0c + a_0^\dagger c^\dagger). &(A5)\cr}
$$
Hereafter we measure energies in units of $J$.

Since the Hamiltonian is composed only of the terms quadratic in the boson
operators, we can solve the problem by using a Green's function
formalism.\[4,5] We consider the following functions,
$$
\eqalignno{
 G(i,i';t) &= -i<T(a_i(t)a_{i'}^\dagger(0))>, \quad
 D(t) = -i<T(c^\dagger(t)c(0))>, &(A6)\cr
 F(i;t) &= -i<T(a_i(t)c(0))>, \quad
 \tilde F(i;t) = -i<T(c^\dagger(t)a_{i}^\dagger(0))>. &(A7)\cr}
$$
Here $T$ is the time-ordering operator.
We take $H_c-J'S^2+J'Sc^\dagger c$ as the unperturbed
Hamiltonian, and $J'S(a_0^\dagger a_0 + a_0 c + a_0^\dagger c^\dagger)$ as
a perturbation.
Then the temporal Fourier transform of the unperturbed Green's functions leads
to
$$
\eqalignno{
 G^0(i,i';\omega) &= {2\over N}\sum_k{\rm e}^{ik(i-i')}
 \left({{\ell_k^2}\over{\omega-E_k+i\delta}}-
       {{m_k^2}\over{\omega+E_k-i\delta}}\right), &(A8)\cr
 D^0(\omega)&={{-1}\over{\omega+J'S-i\delta}},&(A9)\cr}
$$
where
$$
\eqalignno{
 E_k &= 2S(1-\alpha^2\cos^2k)^{1/2}, &(A10)\cr
 \ell_k &=\biggl({{2S+E_k}\over{2E_k}}\biggr)^{1/2},\quad
    m_k =-\biggl({{2S-E_k}\over{2E_k}}\biggr)^{1/2}, &(A11)\cr}
$$
with $-\pi/2<k\leq\pi/2$. The excitation energy $E_k$ has a gap.

Summing up the diagrams to infinite order, the ones of lowest order
shown in Fig.15, we obtain
$$
\eqalignno{
 G(i,i';\omega) &= G^0(i,i';\omega) + G^0(i,0;\omega)
 {{\pi(\omega)}\over{1-G^0(0,0;\omega)\pi(\omega)}} G^0(0,i';\omega),
 &(A12)\cr
 D(\omega) &= D^0(\omega) + D^0(\omega)(J'S)
 {{G^0(0,0;\omega)}\over{1-G^0(0,0;\omega)\pi(\omega)}}(J'S)
 D^0(\omega), &(A13)\cr
 F(i;\omega) &= G^0(i,0;\omega)
 {{1}\over{1-G^0(0,0;\omega)\pi(\omega)}}(J'S)D^0(\omega), &(A14)\cr
 \tilde F(i;\omega) &= D^0(\omega)J'S
 {{1}\over{1-G^0(0,0;\omega)\pi(\omega)}}G^0(0,i;\omega), &(A15)\cr}
$$
where
$$
 \pi(\omega) \equiv J'S + (J'S)^2D^0(\omega). \eqno(A16)
$$
The energy of the bound state $E_B$ is determined by the relation
$$
  1-G^0(0,0;E_B)\pi(E_B)=0. \eqno(A17)
$$
For $J'<2$, one of the bound states is positioned in the region
$-2S\sqrt{1-\alpha^2}<E_B<0$, and another has in $2S<E_B$.
Using the Green's functions, we obtain the averages of spin operators
from the relations,
$$
\eqalignno{
 <S_{0f}^z> &= -S + i\lim_{t\to 0^+}\int_{C_-}D(\omega)
  {\rm e}^{-i\omega t}{{{\rm d}\omega}\over{2\pi}}, &(A18)\cr
 <S_{0c}^z> &= S - i\lim_{t\to 0^-}\int_{C_+}G(0,0;\omega)
  {\rm e}^{-i\omega t}{{{\rm d}\omega}\over{2\pi}}, &(A19)\cr
 <S_{ic}^z> &= S - i\lim_{t\to 0^-}\int_{C_+}G(i,i;\omega)
  {\rm e}^{-i\omega t}{{{\rm d}\omega}\over{2\pi}}. &(A20)\cr}
$$
Here $C_+$ ($C_-$) indicates that the integration path for ${\rm d}\omega$
is taken along the half circle in the upper (lower) plane for complex
$\omega$.

Figure 16 shows the calculated averages of the spin moments
for $J'=0.5$, $S=1/2$.
The sublattice magnetization $<S_{ic}^z>$ (for $i$ far from the impurity
site) decreases with increasing values of $\alpha$,
due to the increase of the quantum fluctuation (curve c).
The spin-wave expansion fails for $\alpha$ close to 1;
for $\alpha\geq 0.98$, $<S_{ic}^z>$ becomes negative.
The average of the spin in the chain closest to the impurity site also
decreases
with increasing values of $\alpha$ (curve b). The values are smaller than
the sublattice magnetization, since the coupling to the impurity spin adds
to fluctuations.
The absolute value of the average of the impurity spin $|<S_{0f}^z>|$
is reduced from the classical value $0.5$,
due to transverse spin fluctuations caused by the coupling to spins
in the chain. They change rather slowly with changing $\alpha$,
except when $\alpha\approx 1$.

The transverse spin correlation between the impurity spin
and the nearest spin in the chain are calculated from
$$
 <S_{0f}^xS_{0c}^x> = i{S\over 2}\lim_{t\to 0^-}\int_{C_+}
 [F(0;\omega)+\tilde F(0;\omega)]
  {\rm e}^{-i\omega t}{{{\rm d}\omega}\over{2\pi}}. \eqno(A21)
$$
Figure 17 shows the values for $J'=0.5$, $S=1/2$.
They are close to the minimum -0.25 when $\alpha\approx 1$.
\vfil \eject
\def\vol(#1,#2,#3){{\bf #1}, #3 (#2)}
\centerline{\bf References}
\item{$^1$}For earlier works, see, {\it Magnetism V}, eds G. T. Rado and
H. Suhl (Academic Press, New York and London, 1973).
\item{$^2$}N. Andrei, K. Furuya, and J. H. Lowenstein,
Rev. Mod. Phys. \vol(55,1983,331).
\item{$^3$}T. Schork and P. Fulde, Phys. Rev. \vol(B50,1994,1345).
\item{$^4$}T. Tonegawa, Prog. Theor. Phys. \vol(40,1968,1195).
\item{$^5$}N. Bulut, D. Hone, D. J. Scalapino, and E. Y. Loh,
Phys. Rev. Lett. \vol(62,1989,2192).
\item{$^6$}T. Tonegawa, M. Kaburagi, N. Ichikawa, and I. Harada,
J. Phys. Soc. Jpn. \vol(61,1992,2890).
\item{$^7$}
Assume that the conduction band is described by the half-filled
one-dimensional Hubbard model:
$H_c=-t\sum_{i\sigma}(c_{i\sigma}^\dagger c_{i+1\sigma}+{\rm H.c.})
 + U\sum_i c_{i\uparrow}^\dagger c_{i\downarrow}^\dagger
 c_{i\downarrow} c_{i\uparrow}$. The conventional Kondo model
corresponds to $U=0$ while here $U/t$ is assumed to be large.
\item{$^8$}B. A. Jones, C. M. Varma, and J. W. Wilkins,
Phys. Rev. Lett. \vol(61,1988,125); Phys. Rev. \vol(B40,1989,324).
\item{$^9$}O. Sakai, Y. Shimizu, and T. Kasuya, Solid State Commun.
\vol(75,1990,81); O. Sakai and Y. Shimizu, J. Phys. Soc. Jpn.
\vol(61,1992,2333); \vol(61,1992,2348).
\item{$^{10}$}R. M. Fye and J. E. Hirsch, Phys. Rev. \vol(B40,1989,4780);
R. M. Fye, Phys. Rev. Lett. \vol(72,1994,916), and references therein.
\item{$^{11}$}H. Tsunetsugu, Y. Hatsugai, K. Ueda, and M. Sigrist,
Phys. Rev. \vol(B46,1992,3175).
\item{$^{12}$}Z. Wang, X.-P. Li, and D.-H. Lee, Phys. Rev.
\vol(B47,1993,11935).
\item{$^{13}$}C. C. Yu and S. R. White, Phys. Rev. Lett.
\vol(71,1993,3886), and references therein.
\item{$^{14}$}T. Barnes, E. Dagotto, J. Riera, and E. S. Swanson,
Phys. Rev. \vol(B47,1993,3196).
\item{$^{15}$}F. D. M. Haldane, Phys. Lett. \vol(93A,1983,464);
Phys. Rev. Lett. \vol(50,1983,1153).
\item{$^{16}$}K. Hida, Phys. Rev. \vol(B45,1992,2207).
\item{$^{17}$}For example, J. S\'olyom and J. Timonen, Phys. Rev.
 \vol(B39,1989,7003).
\item{$^{18}$}We employ a computer program package KOBEPACK/S,
which uses a new coding technique `{\it subspace coding}' for saving
the memory space and CPU time. See, for details,
M. Kaburagi, T. Tonegawa, and T. Nishino, in
{\it Computational Approaches in Condensed Matter Physics}, edited by
S. Miyashita, M. Imada, and T. Takayama, p.179 (Springer, Berlin, 1992).
\item{$^{19}$}Assuming a N-dependence of $\Delta E_1(J',N)$ as
$$
 \Delta E_1(J',N) = \Delta E_1(J') + {{a_1}\over N}
 + {{a_2}\over{N^2}} + \cdots + {{a_p}\over{N^p}},
$$
with $p=3$, 4, $\cdots$, 9, we make a least-squares fit to the finite cluster
results for $N=N_{\rm max}, N_{\rm max}-2, \cdots, N_{\rm max}-2m$
with $m=p$, $m=p+1$, $\cdots$, 9, where $N_{\rm max}=23$ for odd-$N$ series
and $N_{\rm max}=22$ for even-$N$ series. The value of $\Delta E_1(J')$
thus obtained depends on $m$ as well as on $p$.
Averaging these results, we estimate the limiting value $\Delta E_1(J')$
for odd and even values of $N$.
\item{$^{20}$}K. A. Hallberg and C. A. Balseiro, unpublished.
\item{$^{21}$}T. Sakai and M. Takahashi, Phys. Rev. \vol(B42,1990,1090);
K. Hida, J. Phys. Soc. Jpn. \vol(60,1991,1347);
S. R. White and D. A. Huse, Phys. Rev. \vol(B48,1993,3844).
\item{$^{22}$}J. Kondo, Phys. Rev. \vol(154,1967,644).
\item{$^{23}$}D. G. Clarke, T. Giamarchi, and B. I. Shraiman,
Phys. Rev. \vol(B48,1993,7070).
\vfil \eject
\centerline{Figure Captions}
\parindent 3pc
\item{Fig. 1.}
Sketches of system with (a) a single impurity, (b) two impurities,
and (c) impurities forming a lattice.
Open circles represent spins.
Furthermore $J$ and $J'$ represent the exchange interactions between spins
in the chain and  between an impurity spin and a spin in the chain,
respectively.
\item{Fig. 2.}
Binding energy $\Delta E_1(J';N)$ for systems with a single impurity.
The solid lines correspond to even numbers N, from $N=4$ (lowest line) to
$N=22$ (upmost line).
The dashed lines correspond to odd numbers N, from $N=5$ (upmost line) to
$N=23$
(lowest line). The values extrapolated to $N\to\infty$ are shown
by open circles with error bars as indicated.
The broken line represents ${3\over 4}J'$, i.e.,
the energy of the  perfect singlet.
\item{Fig. 3.}
Correlation function $-<{\bf S}_{0f}\cdot{\bf S}_{0c}>$
for systems with a single impurity.
Solid and dashed lines correspond to even and odd numbers of N,
in the same way as in Fig.2.
The extrapolated values are shown by open circles with error bars as indicated.
\item{Fig. 4.}
Binding energy per impurity spin $\Delta E_2^{\rm nn}(J',N)/2$
for systems with two impurities placed on nearest-neighbor sites.
Solid lines correspond to even numbers N, from $N=4$ (lowest line) to
$N=22$ (upmost line).
Dashed lines correspond to odd numbers N, from $N=3$ (upmost line) to $N=21$
(lowest line).
The extrapolated values are shown by open circles with error bars.
\item{Fig. 5.}
Binding energies per impurity spin, $\Delta E_2^{\rm nn}(J')/2$,
$\Delta E_2^{\rm nnn}(J')/2$, and $E_B^{\rm lat}(J')$,
in comparison with $\Delta E_1(J')$.
\item{Fig. 6.}
Correlation function $-<{\bf S}_{0f}\cdot{\bf S}_{0c}>$
between an impurity spin and the spin in the chain closest to it,
when two impurities are placed on nearest-neighbor sites.
Solid and dashed lines correspond to even and odd numbers of N,
in the same way as in Fig.4.
The extrapolated values are shown by open circles with error bars as indicated.
\item{Fig. 7.}
Correlation function
$-<{\bf S}_{0f}\cdot{\bf S}_{1f}>$ for two impurity spins
placed on nearest-neighbor sites.
Solid lines correspond to even numbers N, from $N=4$ (upmost line) to
$N=22$ (lowest line).
Dashed lines correspond to odd numbers N, from $N=3$ (lowest line) to $N=21$
(upmost line).
The extrapolated values are shown by open circles with error bars as indicated.
\item{Fig. 8.}
Binding energy per impurity spin $E_B^{\rm lat}(J';N)$ for impurity lattices.
Solid lines correspond to even numbers N, from $N=4$ (lowest line) to
$N=14$ (upmost line).
Dashed lines correspond to odd numbers N, from $N=3$ (upmost line) to $N=13$
(lowest line).
The extrapolated values are shown by open circles with error bars as indicated.
\item{Fig. 9.}
Correlation function $-<{\bf S}_{0f}\cdot{\bf S}_{0c}>$ between
one of the impurity spins of the lattice and the nearest spin in the chain.
Solid and dashed lines correspond to even and odd numbers of N,
in the same way as in Fig. 8.
\item{Fig. 10.}
Correlation function $-<{\bf S}_{0f}\cdot{\bf S}_{1f}>$
between a pair of impurity spins on nearest-neighbor sites,
for impurity lattices.
Solid lines correspond to even numbers of N, from $N=4$ (upmost line) to
$N=14$ (lowest line).
Dashed lines correspond to odd numbers N.
The extrapolated values are shown by open circles with error bars as indicated.
\item{Fig. 11.}
Energy gap $\Delta(J',N)$ for impurity lattices.
Solid lines correspond to even numbers of N, from $N=4$ (upmost line) to
$N=14$ (lowest line).
Dashed lines correspond to odd numbers of N, from $N=3$ (upmost line) to $N=13$
(lowest line).
The extrapolated values are shown by open circles with error bars as indicated.
The broken line represents $J'-{1\over 2}$, which is the value of
$\Delta(J')$ valid for $J'>>1$.
\item{Fig. 12.}
Binding energy per impurity spin $E_B^{\rm lat}(J';N)$ with $J'<0$
for impurity lattices.
Solid and dashed lines correspond to even and odd numbers of N,
in the same way as in Fig.8.
The curve for N=3 is identical to the broken line ${1\over 4}|J'|$, i.e.,
the energy of the perfect triplet.
The extrapolated values are shown by open circles with error bars as indicated.
\item{Fig. 13.}
Correlation function
$-<{\bf S}_{0f}\cdot{\bf S}_{0c}>$ between one of the impurity
spins of the lattice and the nearest spin in the chain, with $J'<0$.
Solid and dashed lines correspond to even and odd numbers of N,
in the same way as in Fig.8.
\item{Fig. 14.}
Energy gap $\Delta(J',N)$ with $J'<0$ for impurity lattices.
Solid and dashed lines correspond to even and odd numbers of N,
in the same way as in Fig.11.
The extrapolated values are shown by open circles with error bars as indicated.
The broken line represents the Haldane gap for $S=1$.
\item{Fig. 15.}
Diagrams for the Green's functions. Solid lines represent
$G^0(i,i';\omega)$, and broken lines represent $D^0(\omega)$.
Crosses represent the interaction $J'S$ at site $i=0$.
\item{Fig. 16.}
Average of the impurity spin $-<S_{0f}^z>$, of the spin in the chain closest to
it $<S_{0c}^z>$, and of a spin in the chain far away from it $<S_{ic}^z>$,
as a function of $\alpha$ with $J'=0.5$.
\item{Fig. 17.}
Transverse spin correlation function $<S_{0f}^xS_{0c}^x>$ as a function of
$\alpha$ with $J'=0.5$.
\bye